\documentclass[epj]{webofc}
\usepackage[varg]{txfonts}   
\woctitle{Physics at the Magnetospheric Boundary}
\def\sqiglt{\hbox{\rlap{\lower.55ex \hbox {$\sim$}}
	\kern-.3em \raise.4ex \hbox{$<$}\,}}
\def\etal{et\,al.}
\begin{document}
\title{The Magnetospheric Boundary in Cataclysmic Variables}
%
%

\author{Coel Hellier}


\institute{Astrophysics Group, Keele University, Keele, U.K.}

\abstract{%
The magnetic cataclysmic variables (MCVs) present a wealth of observational diagnostics for studying accretion flows interacting with a magnetosphere.  Spin-period pulsations from the rotation of the white dwarf are seen in optical light, in the UV and X-ray bands, and in polarimetry, and modelling these can constrain the size and location of the accretion footprints on the white-dwarf surface.  Tracing these back along field lines can tell us about the transition region between the stream or disk and the magnetosphere.   Further, optical emission lines give us velocity information, while analysis of eclipses gives spatial information. 

I discuss MCVs (particularly FO~Aqr, V405~Aur, XY~Ari and EX~Hya, but also mentioning PQ~Gem, GK~Per, V2400~Oph, HT~Cam, TX~Col, AO~Psc, AE~Aqr, WZ~Sge, V1223~Sgr and DQ~Her), reviewing what observations tell us about the disk--magnetosphere boundary.   The spin-period variations are caused by a mixture of geometric effects and absorption by the accretion flow, and appear to show that the accretion disk feeds onto field lines differently in different systems, being sometimes along field lines ahead of the magnetic pole and sometimes behind the pole.  

During outbursts, when the accretion flow increases by orders of magnitude, the disk pushes the magnetosphere inwards, and appears to feed field lines over a much greater range of magnetic azimuth.  The non-equilibrium outburst behaviour shows an even richer phenomenology than in quiescence, adding DNOs and QPOs into the mix.
}
\maketitle
\section{Why study cataclysmic variables?}
\label{intro}
The magnetic cataclysmic variables -- interacting binary stars with accretion onto a white dwarf -- present one of the most advantageous opportunities for studying the interaction of an accretion flow with a magnetosphere. The reasons include: (1) the magnetic field of a white dwarf is frozen and thus stable, instead of being a changeable, dynamo-driven field, as in young stars.  (2) the magnetic and gravitational field strengths conspire such that light from magnetically controlled accretion can be prominent from the infra-red and optical to the UV and X-ray (Fig.~1). (3) The orbital periods of the binaries (hours) and the spin periods of the white dwarfs ($\sim$ 10 min) result in variations on easy-to-observe timescales.  (4) Some systems are eclipsing, giving valuable geometric probes of the location of accretion light. (5) The occurrence of semi-regular outbursts (in which the accretion flow can jump by orders of magnitude) enables us to study non-equilibrium behaviour.  

\begin{figure}
\centering
\includegraphics[width=14cm]{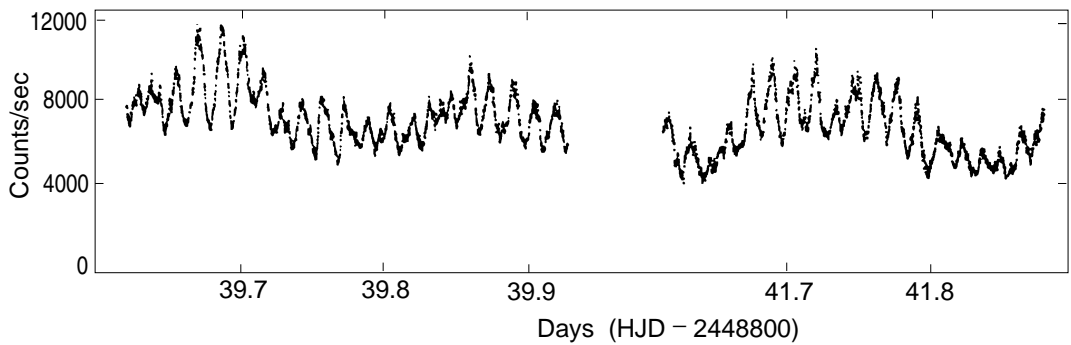}
\caption{FO~Aqr shows prominent variations at both the 20.9-min white-dwarf spin period and the 4.85-hr orbital period; optical photometry by Patterson \etal\ (1998).}
\label{chfig-1}       
\end{figure}

In this review I'll consider what the observations tell us about the magnetospheric boundary, and how the accretion flow from the secondary star comes to be magnetically controlled (see the companion review by Dayal Wickramasinghe for a theoretical perspective, while for an introduction to cataclysmic variables see Hellier 2001 or the very comprehensive Warner 1995).    

Although I mention the phase-locked AM Her sub-class (`polars'), I'll focus mainly on the systems containing an accretion disk (`intermediate polars' or IPs)  and on the boundary between the accretion disk and the magnetosphere.  

\section{Streams in the highest-field systems}
When the white dwarf's magnetic field is above $\sim$\,10 MG it usually prevents the accretion stream from orbiting the white dwarf and forming itself into a disk.  It can also lock the white-dwarf rotation to the orbital cycle, thus producing an AM Her star. 

The track of the accretion flow can be deduced from: (1) modelling X-ray lightcurves, in terms of accretion footprints passing over the white-dwarf limb, plus absorption by the stream occulting the footprints; (2) modelling polarimetry of the cyclotron emission produced by the stream near the footprint; (3) analysing emission-line Doppler shifts; and (4) using eclipses to locate the accretion light  (e.g.\  Buckley \etal\ 2000,  G\"ansicke \etal\ 2001; Schwope \etal\ 2000, 2001, 2003;  Potter \etal\ 2004;  Rodrigues \etal\ 2006; Campbell \etal\ 2008;  Thomas  \etal\ 2012; Silva \etal\ 2013). 

Such studies show that the stream usually follows a ballistic path in the outer Roche lobe, then has an extended transition region as it encounters the magnetic field, and then becomes magnetically controlled as it follows field lines onto the white dwarf.   Since the transition region is usually extended, material links to a range of field lines, and can end up accreting at both magnetic poles.   Further, since the degree to which the stream punches into the magnetosphere will depend on the accretion rate, the accretion geometry, and whether one or two poles are being fed, can depend on the mass-transfer rate, which can vary by orders of magnitude.

\section{Transitional pole-flipping systems}
A small number of AM Her stars have slightly different spin and orbital periods (perhaps being knocked out of synchronism by nova explosions). The orientation of the field with respect to the stream then changes on the beat cycle between the spin and orbital periods ($1/P_{\rm beat} = 1/P_{\rm spin} - 1/P_{\rm orb}$).  Thus we see the stream flipping from accreting predominantly onto one pole to accreting predominantly onto the other (on a timescale of $\sim$ 10 d). As is perhaps expected, the transition can be messy rather than abrupt and thus the accretion flow can be much more spread out than in other AM Her stars (e.g.\ Schwarz \etal\ 2005). 

While a high magnetic field will tend to lock the white-dwarf spin to the orbit, with a lower field ($\sim$\,1 MG) the white-dwarf is more likely to find an equilibrium where the magnetosphere's spin matches the motion of the accretion stream at the radius at which it would tend to circularise. This produces spin periods of order 0.1\,$P_{\rm orb}$ (King \&\ Lasota 1991), and many IPs are seen to have this ratio (though plenty do not).

V2400 Oph is an important system having a white dwarf spinning at 8\%\ of the orbital period, as revealled by a cyclic variation in polarised light (Buckley \etal\ 1997). Periodicities in X-rays, optical light and emission-line velocities, however, are at the beat period.  An X-ray modulation at the beat period had been previously predicted (e.g.\ Hellier \etal\ 1989) to be a diagnostic of the pole-flipping stream accretion expected in a discless, stream-accreting IP. Buckley \etal\ had thus found the first such system.    Study of the emission-line profiles produces a reasonably good match to a simple model of the expected pole-flipping stream (Hellier \&\ Beardmore 2002). 

\section{Disk-fed magnetic cataclysmic variables}
While asynchronous AM Hers and the diskless IP V2400 Oph show stream-fed accretion, the observational signs are that in most IPs the accretion flow forms into a disk (e.g.\ Hellier 1991). In the `diamagnetic blob' model of Wynn \&\ King (1995) the stream breaks up into blobs that behave semi-ballistically, with magnetic drag caused by the field acting on their surfaces.  If such blobs can survive long enough to circle the white dwarf they can accumulate to produce a disk. 

In such a system we expect to have a near-normal disk far from the white dwarf, a transition region (at a radius set by the field strength and accretion rate), and then accretion down field lines within the magnetosphere.

\subsection{Overflowing stream material} 
Even when most accretion flows through a disk this may not be the whole story, since the stream from the secondary star can be thicker than the outer disk edge, and thus partially overflow the disk.  There is evidence for this in many IPs.  A well-studied example is FO Aqr, where the presence of an intermittent and variable X-ray pulse at the beat period, in addition to the dominant pulse at the spin period, indicates that part of the stream overflows the disk as far as the magnetosphere (e.g.\ Norton \etal\ 1992; Hellier 1993; Beardmore \etal\ 1998). 

TX Col is notable for showing a strong X-ray beat pulse on some occasions and no beat pulse on others (e.g.\ Norton \etal\ 1997),  thus showing that the degree of stream overflow can be very variable.  The evidence for such overflow is not solely from X-ray observations.  In AO Psc, absorption features in UV lines appear to show overflowing-stream material interacting with the magnetosphere, changing direction as the white dwarf rotates (Hellier \&\ van Zyl 2005).  

EX~Hya had been suggested to have an overflowing stream in outburst, based on analysis of emission lines (Hellier \etal\ 1989).  A prediction that it should then show an X-ray beat period in outburst (when it doesn't in quiescence) was later confirmed with {\em RXTE\/} (Hellier \etal\ 2000).

\section{Can we locate the disk--magnetosphere transition?}
To estimate the extent of the magnetosphere one can measure the field strength using polarised light, though this is only possible in a minority of systems (e.g.\ V405 Aur; Piirola \etal\ 2008). 

A second method is to assume an equilibrium between the rotation of the magnetosphere and the Keplerian motion at the inner disk edge, whence the radius follows from the spin period and the white-dwarf mass.  Whether IPs are in equilibrium, however, is questionable.  FO Aqr is near the expected equilibrium of $P_{\rm spin} = 0.1 P_{\rm orb}$ and shows episodes of the white dwarf spinning up, and of spinning down (e.g.\ Williams 2003),  suggesting that it fluctuates about an equilibrium. 

AE Aqr, on the other hand, has a very short spin period of 33 s and shows long-term spin-down (de Jager \etal\ 1994), suggesting that it is far from equilibrium and indeed magnetically propelling material outwards (e.g.\ Wynn, King \&\ Horne 1997; Oruru \&\ Meintjes 2012). At the other extreme EX~Hya has a very long spin period of 67 min. 

\begin{figure}
\centering
\includegraphics[width=15cm]{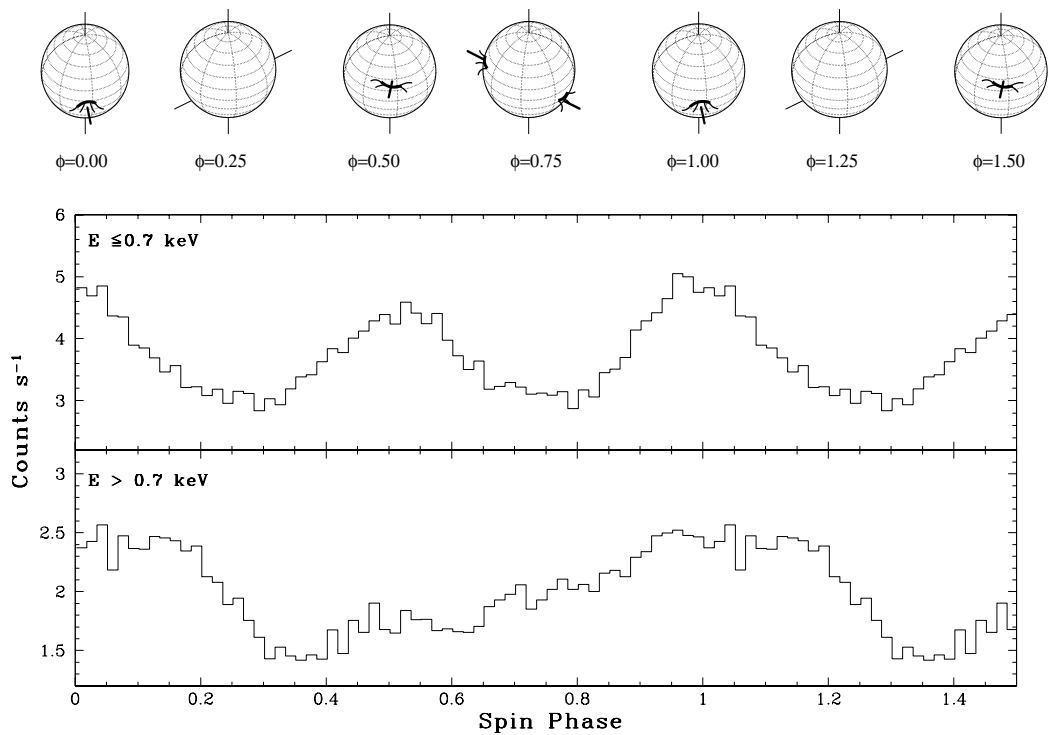}
\caption{The accretion footprints in V405~Aur deduced from the X-ray spin pulse (Evans \&\ Hellier 2004).}
\label{chfig-2}       
\end{figure}

\subsection{The case of EX Hya}
If EX Hya's white dwarf is spinning near equilibrium then the magnetosphere must nearly fill the white dwarf's Roche lobe, and there would be only a  narrow `accretion ring' instead of a full disc (e.g.\ King \&\ Wynn 1999; Belle \etal\ 2002; Mhlahlo \etal\ 2007). 

However, EX Hya is a grazing eclipser, with the lower part of the white dwarf and material falling to that pole being eclipsed, while the upper pole remains un-eclipsed (as demonstrated by partial X-ray eclipses; e.g.\ Hoogerwerf, Brickhouse \&\ Mauche 2005).  Thus by analysing eclipse timings as a function of spin-cycle phase one can locate the centroid of the spin-varying emission.  This gives a centroid at only $\sim$\,2 white-dwarf radii from the white-dwarf centre (Siegel \etal\ 1989), and thus points to a magnetospheric radius of only $\sim$\,4 white-dwarf radii.    

This is also in line with estimates from the velocities of the emission-line components that don't vary with the spin period (and thus are likely from the disk; e.g.\ Hellier \etal\ 1987), and with the fact that EX Hya shows no polarisation, which is unlikely if the magnetosphere dominates the primary's Roche lobe.   Thus, the observations point to EX Hya having a small magnetosphere and being far from equilibrium. The spin period also shows a long-term secular decrease (e.g.\ Hellier \&\ Sproats 1992), in keeping with this suggestion.

Some observers have preferred to argue for a large magnetosphere, based on the finding of variations on the spin cycle in the low-velocity core of the emission lines  (Mhlahlo \etal\ 2007). However, since cyclical motion will cross zero radial velocity twice per cycle, this is not a sufficient argument.    Further, arguments based on analysing the orbital cycle as a function of spin phase (e.g.\ Belle \etal\ 2002) are suspect since in EX~Hya the two periodicities are commensurate (the spin period being very nearly 2/3rds of the orbital period) and thus they cannot readily be separated, with folds on one periodicity being contaminated by variations on the other.  

\begin{figure}
\centering
\includegraphics[width=6cm]{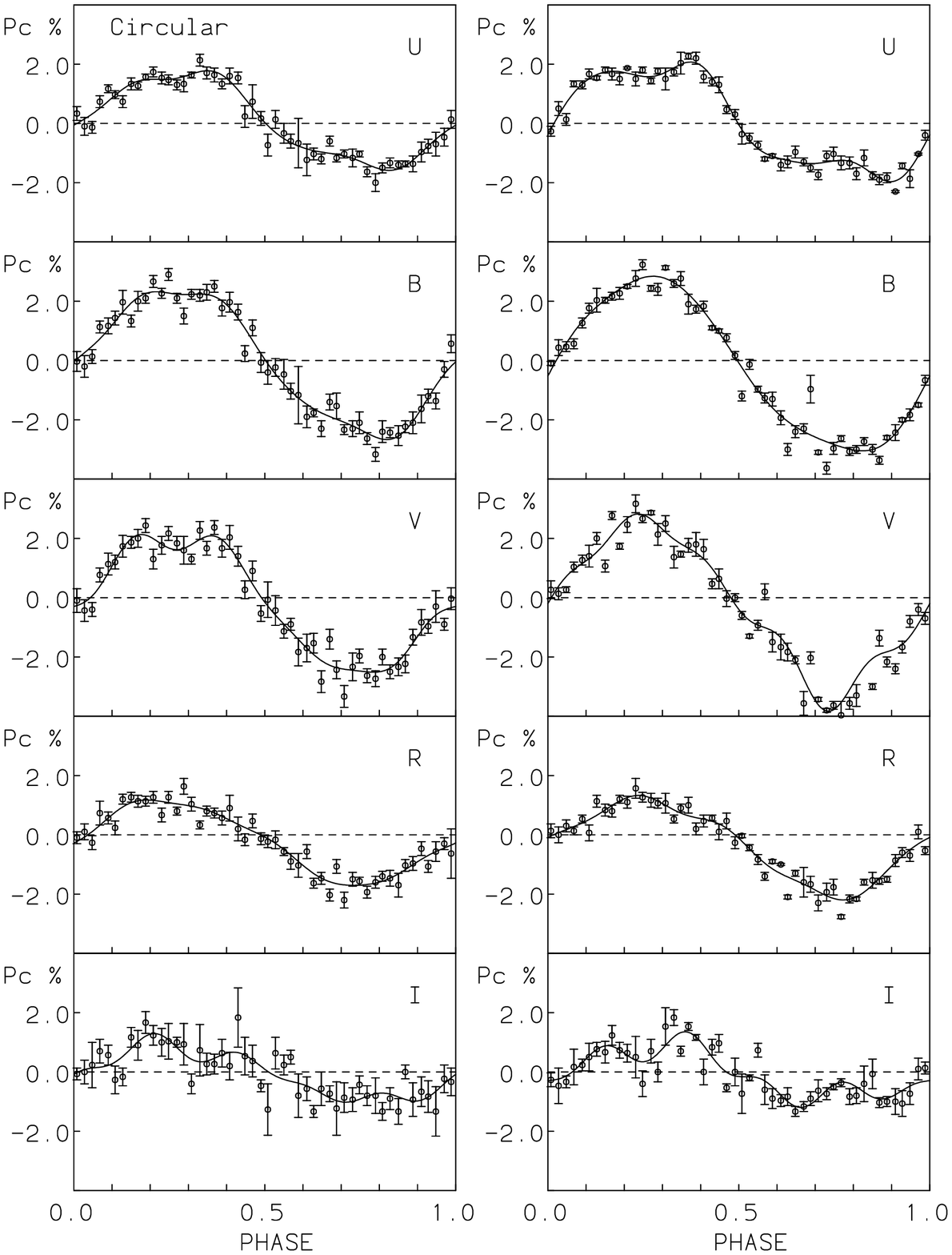}\includegraphics[width=8cm]{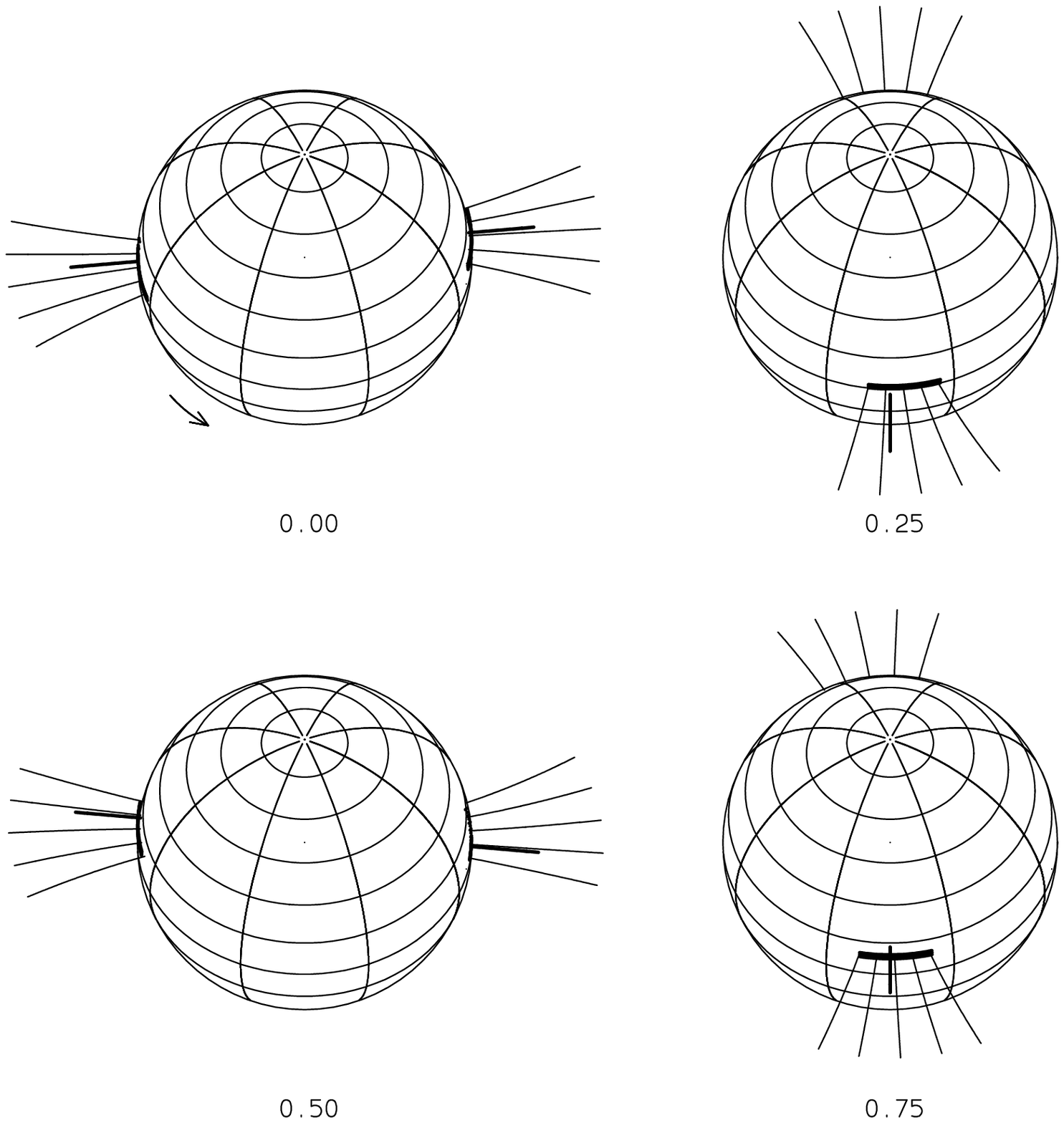}
\caption{The accretion footprints in V405~Aur deduced from polarimetry (Piirola \etal\ 2008), are compatible with those from the X-ray study in Fig.~2.}
\label{chfig-3}       
\end{figure}

\section{What is the size and shape of the feeding region?}
Deducing the  size and shape of the region over which material feeds into field lines is hard. The usual way is to analyse the spin-pulse profiles (in the optical and X-ray and in polarised light) and interpret them in terms of accretion footprints on the white dwarf and obscuration by accretion curtains sweeping around; one can then trace back along field lines to a deduced feeding region. 

A similar method is to use eclipses to locate X-ray-emitting regions on the white dwarf and trace these back to the feeding region.    Few systems, though, show both strong X-ray flux and eclipses, of which a leading example is XY~Ari.  Analysis of eclipses in XY~Ari shows that the X-ray egresses are short compared to the white-dwarf egress, such that the footprint covers only $\sim$\,0.002 of the white-dwarf surface (Hellier 1997). This result, though, comes from adding up many different egresses to attain sufficient photons.  What we really need is an X-ray satellite capable of analysing individual eclipses.  Taken at face value the XY~Ari result implies a feeding region restricted in azimuth (\sqiglt 30$^{\circ}$).

Corroboration of this comes from other estimates of the size of the accretion footprint in IPs. For example  V405 Aur shows a soft, blackbody X-ray component, presumably from the white-dwarf surface near the accretion footprints that is heated by harder X-ray emission. This enables an estimate of the size of the heated area of order 0.001 (Haberl \&\ Motch 1995; Evans \&\ Hellier 2004).   Similarly, Hayashi \etal\ (2011)  deduce a footprint of  $<$\,0.007 of the white dwarf in V1223~Sgr, based on analysing the X-ray spectrum.

Modelling of the spin pulse in V405 Aur in terms of accretion footprints shows a fair degree of correspondence between studies using X-rays (Fig.~2; Evans \&\ Hellier 2004) and studies using optical polarimetry (Fig.~3; Piirola \etal\ 2008), giving some confidence in the validity of the conclusions.  

Another important case is HT Cam, which has an energy-independent X-ray spin pulse, suggesting that absorption effects are unimportant, and thus that the accretion curtains never obscure the accretion footprints.  The X-ray pulse can then be modelled as a simple geometrical effect of footprints appearing or disappearing over the white-dwarf limb (Evans \&\ Hellier 2005). The best models  have a footprint arc extending over 20 or 30 degrees in magnetic azimuth (with the X-ray brightness assumed to be weighted to the centre of the arc by a $\cos^{2}\theta$ dependence). Simplistically, this would imply a similar extent of the feeding region at the magnetosphere.    Most other IPs have X-ray spin-pulses showing such complex absorption effects that similar modelling is too ambiguous.  

\begin{figure}
\centering
\includegraphics[width=12cm]{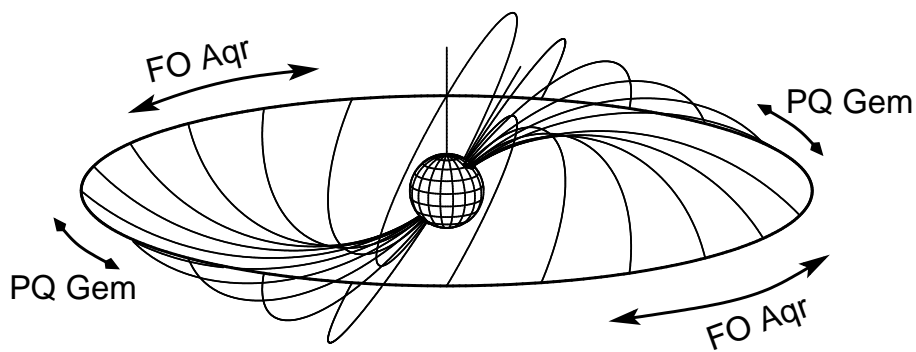}
\caption{The feeding in FO Aqr appears to be behind the magnetic pole whereas that in PQ~Gem appears to precede the pole (drawn here assuming counter-clockwise rotation).} 
\label{chfig-4}       
\end{figure}

\subsection{Is feeding ahead of or behind the pole?}
One can ask whether the field lines picking up material from the disk are aligned with the magnetic axis, or whether they are predominantly behind or ahead of the axis.  One can also ask whether this is related to how fast the magnetosphere is spinning relative to the disk's inner edge, and thus, perhaps,  to whether the white dwarf is being spun up or down.  

Two interesting systems are FO Aqr and PQ Gem (Fig.~4).  In FO Aqr a prominent X-ray absorption dip occurs at a phase implying that the infalling material is trailing the magnetic pole, by up to  a quarter of a cycle (Evans \etal\ 2004). FO~Aqr's white dwarf has been seen to be spinning down, and more recently to be spinning up (e.g.\ Williams 2003), but there is no clear indication that the phasing of spin-pulse features changes markedly between the two epochs (e.g.\ Beardmore \etal\ 2008). 

In PQ Gem the infalling material appears to be ahead of the magnetic pole, by 0.1 in phase (Evans, Hellier \&\ Ramsay 2006),  the opposite to FO Aqr.  PQ Gem's white dwarf shows a long-term spin down (Mason 1997; Evans \etal\ 2006).   Overall there seems to be no straightforward interpretation of these results, and we need similar studies of more systems to look for patterns.

\section{How are things changed by outbursts?} 
During a disk-instability outburst the accretion rate pushing at the magnetosphere climbs by orders of magnitude.  Although only semi-predictable, outbursts are valuable for studying non-equilibrium behaviour of the disk--magnetosphere boundary.   I highlight here two examples. 

XY~Ari was caught in outburst by the X-ray satellite {\sl RXTE}.  Near the peak of the outburst an eclipse egress occurred earlier than expected (Hellier \etal\ 1997).  The likeliest interpretation is that the accretion footprints were then much larger, covering more of the white dwarf, and thus that material was feeding onto field lines over a much greater range of magnetic azimuth (Fig.~5). 

A suggestion of the same comes from GK~Per in outburst, also monitored by {\sl RXTE}.    Analysis of the outburst spin pulsation again suggests that accretion was flowing from all azimuths onto a much larger swath of the white dwarf. Also, with the disk having pushed inwards, the accretion curtain obscured the footprints, producing a deep absorption dip and a much larger spin-pulse modulation than in quiescence (Hellier, Harmer \&\ Beardmore 2004; Vrielmann, Ness \&\ Schmitt 2005). 

\begin{figure}
\centering
\includegraphics[width=5.5cm]{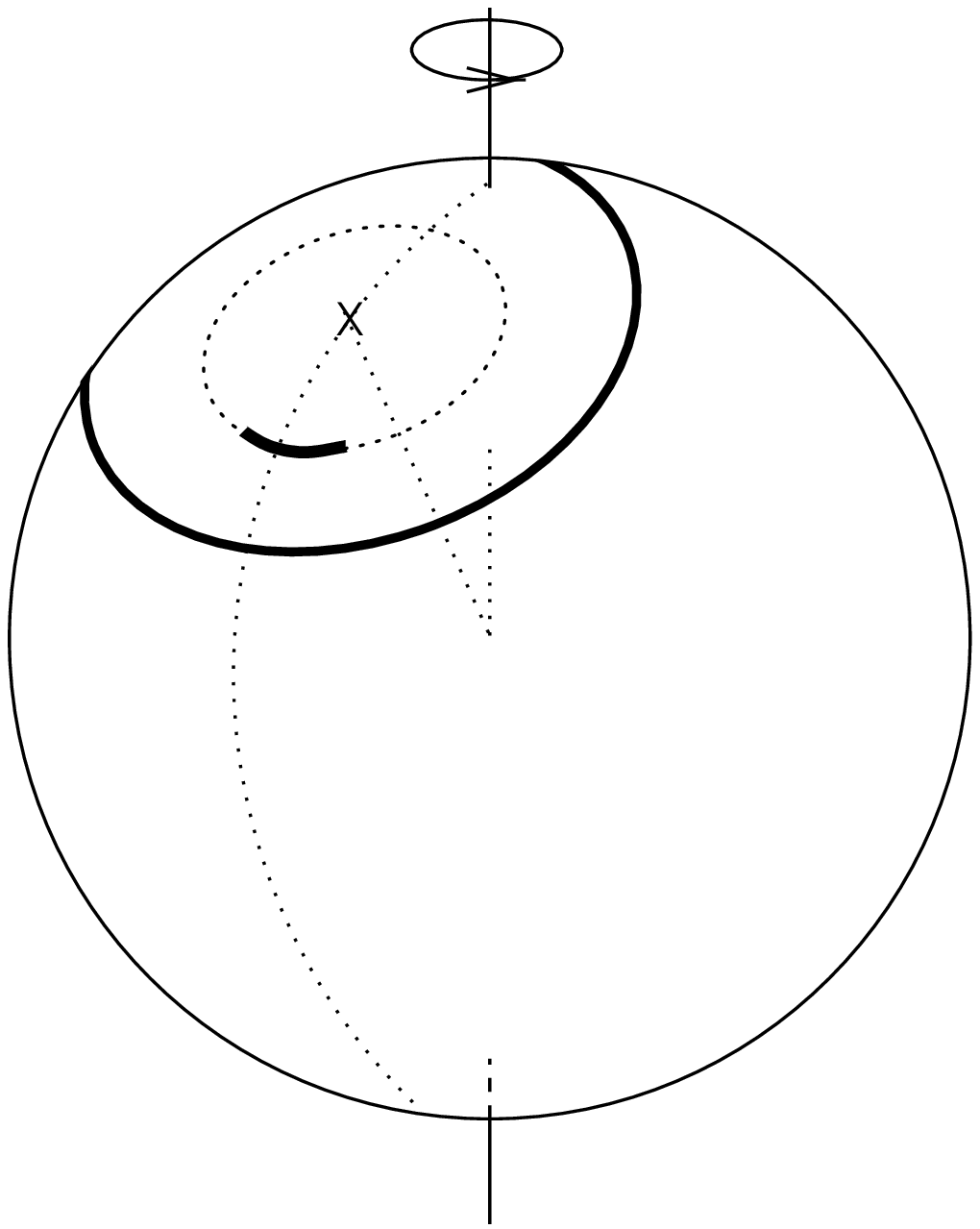}\parbox[b]{5cm}{Fig.~5: A schematic of how the accretion footprints in XY~Ari change between quiescence (inner, shorter, bold line) and outburst (outer ring).}
\label{chfig-5}       
\end{figure}

\subsection{QPOs/DNOs}
One might expect the outburst behaviour of a disk--magnetospheric interaction to be particularly rich in phenomena, and this is borne out. The literature on temporary, non-coherent oscillations in CV outbursts is vast, though models developed in a series of papers primarily by Warner \& Woudt are making sense of it.   For details see the review by Warner (2004) and the later papers Warner \& Woudt (2006) and Woudt \&\ Warner (2009). 

A brief summary is that in the outburst of a non-magnetic CV the disk material seems to spin up a central belt of the white dwarf, and the resulting dynamo generates a magnetic field and a temporary magnetosphere.  Further, this appears to excite slowly moving prograde waves of material at the disk--magnetosphere boundary.  Semi-coherent photometric modulations are then seen at (1) the period of the spinning central belt and magnetosphere (`dwarf nova oscillations' or DNOs at $\sim$ 10 s); (2) the period of the slower-spinning white-dwarf body (`long-period' DNOs at typically 4 times the DNO period); (3) the period of the prograde travelling waves (QPOs, at typically 15 times the DNO period); and (4) the beat-frequency between the spinning magnetosphere and the travelling waves. 

GK~Per has a magnetic field even in quiescence, but in outburst it shows  QPOs (Watson, King \&\ Osborne 1985) that appear to be similar to those seen in outbursts of non-magnetic CVs, including having the same 15:1 period ratio. Thus these QPOs might also result from prograde travelling waves at the inner disk boundary.     An interesting question is whether such phenomena occur in IPs in quiescence (perhaps being overlooked owing to the longer timescales and faintness), or whether they are only excited by outbursts. 

WZ~Sge is another star showing outbursts and combining properties of non-magnetic and magnetic systems.  A possible white-dwarf spin period of 28.9 s is even faster than in the propeller system AE~Aqr, and it may have a field on the verge of qualifying it for IP status (e.g.\ Warner \&\ Pretorius 2008). Thus WZ~Sge and GK~Per are important for unifying an understanding of outbursts in magnetic and non-magnetic CVs.     In that regard, note also the observation of spiral density waves in the disk of the IP DQ~Her (Bloemen \etal\ 2010), something previously seen in dwarf-novae outbursts.

\section{Concluding remarks} 
New developments include new eclipsing IPs, such as IPHAS\,J062746.41+014811.3 (Aungwerojwit \etal\ 2012) and (the sadly faint) V597 Pup (Warner \& Woudt 2009), which could lead to valuable detailed studies to add to those on EX~Hya and XY~Ari.  The newly discovered IP CC Scl (Woudt \etal\ 2012) shows both outbursts and superhumps, and is thus more similar to the non-magnetic dwarf novae than most IPs.  Woudt \etal\ suggest that it could have a relatively small magnetic moment, which thus gives us another valuable system for studying the range of scenarios in MCVs. 

 The puzzling systems FS~Aur seems to be now confirmed as an IP with a precessing white dwarf (e.g.\ Neustroev \etal\ 2013), adding a new type of phenomenology to an already complex topic.  Also notable is the turning of 8-m-class telescopes onto old-favourite stars, obtaining unprecedented data quality (e.g.\ Beuermann \& Reinsch 2008), and the use of new fast-readout detectors (e.g.\ Bloemen \etal\ 2013).   

Lastly, X-ray sky surveys are leading to boom times in the discovery of IPs (e.g.\ Patterson \etal\ 2011; Scaringi \etal\ 2011; Bernardini \etal\ 2012; Kozhevnikov 2012;  Middleton \etal\ 2012; Masetti, Nucita \&\ Parisi 2012; Girish \& Singh 2012; Bernardini \etal\ 2013), most of which are only beginning to be assimilated into the literature.  Hence we can expect major ongoing developments in our understanding of MCVs.


\begin{thebibliography}{}

\bibitem{1}
Aungwerojwit, A., G\"ansicke, B. T., Wheatley, P. J., Pyrzas, S., Staels, B., Krajci, T., Rodr\'iguez-Gil, P., ApJ, \textbf{758}, 79 (2012) 

\bibitem{2}
Beardmore, A. P., Mukai, K., Norton, A. J., Osborne, J. P., Hellier, C., MNRAS, \textbf{297}, 337 (1998) 

\bibitem{3}
Belle, K. E., Howell, S.B., Sirk, M., Huber, M.E., ApJ, \textbf{577}, 359 (2002)

\bibitem{4}
Bernardini, F., de Martino, D., Falanga, M., Mukai, K., Matt, G., Bonnet-Bidaud, J.-M., Masetti, N., Mouchet, M., A\&A, \textbf{542}, A22 (2012) 

\bibitem{5}
Bernardini, F. \etal\ MNRAS, in press, arXiv:1308.1230 (2013) 

\bibitem{6}
Beuermann, K., Reinsch, K.,  A\&A, \textbf{480}, 199 (2008) 

\bibitem{7}
Bloemen, S., Marsh, T. R., Steeghs, D., \O stensen, R. H., MNRAS, \textbf{407}, 1903 (2010)

\bibitem{8}
Bloemen, S., Steeghs, D., De Smedt, K., Vos, J., G\"ansicke, B. T., Marsh, T. R., Rodriguez-Gil, P., MNRAS, \textbf{429}, 3433 (2013) 

\bibitem{9}
Buckley, D. A. H., Cropper, M, van der Heyden, K., Potter, S. B., Wickramasinghe, D. T., MNRAS, \textbf{318}, 187 (2000) 

\bibitem{10}
Buckley, D.A.H., Haberl, F., Motch, C., Pollard, K., Schwarzenberg-Czerny, A., Sekiguchi, K. MNRAS, \textbf{287}, 117 (1997) 

\bibitem{11}
Campbell, R. K., Harrison, T. E., Kafka, S., ApJ, \textbf{683}, 409 (2008)

\bibitem{12}
de Jager, O.C., Meintjes, P.J., O'Donoghue, D., Robinson, E.L., MNRAS, \textbf{267}, 577 (1994) 

\bibitem{13}
Evans, P. A., Hellier, C., MNRAS, \textbf{353}, 447 (2004) 

\bibitem{14}
Evans, P. A., Hellier, C., MNRAS, \textbf{359}, 1531 (2005) 

\bibitem{15}
Evans, P.A., Hellier, C., Ramsay, G., MNRAS, \textbf{369}, 1229 (2006)

\bibitem{16}
Evans, P.A., Hellier, C., Ramsay, G., Cropper, M., MNRAS, \textbf{349}, 715 (2004)

\bibitem{17}
G\"ansicke, B. T., Fischer, A., Silvotti, R., de Martino, D., A\&A, \textbf{372}, 557 (2001)

\bibitem{18}
Girish, V., Singh, K. P., MNRAS, 427, 458 (2012)

\bibitem{19}
Haberl, F., Motch, C., A\&A, \textbf{297}, L37 (1995)

\bibitem{20}
Hayashi, T., Ishida, M., Terada, Y., Bamba, A., Shionome, T., PASJ, \textbf{63,} S739 (2011)

\bibitem{21}
Hellier, C., MNRAS, \textbf{251}, 693 (1991) 

\bibitem{22}
Hellier, C., MNRAS, \textbf{265}, L35 (1993) 

\bibitem{23}
Hellier, C., MNRAS, \textbf{291}, 71 (1997)

\bibitem{24}
Hellier, C., {\it Cataclysmic variable stars}, (Springer, Berlin 2001)

\bibitem{25}
Hellier, C., Beardmore, A.P., MNRAS, \textbf{331}, 407 (2002)

\bibitem{26}
Hellier, C., Harmer, S., Beardmore, A.P., MNRAS, \textbf{349}, 710 (2004) 

\bibitem{27}
Hellier, C., Kemp, J., Naylor, T., Bateson, F. M., Jones, A., Overbeek, D., Stubbings, R., Mukai, K., MNRAS, \textbf{313}, 703 (2000) 

\bibitem{28}
Hellier, C., Mason, K.O., Rosen, S.R., Cordova, F.A., MNRAS,  \textbf{228}, 463 (1987) 

\bibitem{29}
Hellier, C., Mason, K. O., Smale, A. P., Corbet, R. H. D., O'Donoghue, D., Barrett, P. E., Warner, B., MNRAS, \textbf{238}, 1107 (1989) 

\bibitem{30}
Hellier, C., Mukai, K., Beardmore, A., MNRAS, \textbf{292}, 397 (1997)

\bibitem{31}
Hellier, C., Sproats, L.N., IBVS, \textbf{372}4, 1 (1992)

\bibitem{32}
Hellier, C., van Zyl, L., ApJ, \textbf{626}, 1028 (2005)

\bibitem{33}
Hoogerwerf, R. Brickhouse, N.S., Mauche, C.W., ApJ, \textbf{628}, 946 (2005)

\bibitem{34}
King, A. R., Lasota, J.-P., ApJ, \textbf{378}, 674 (1991) 

\bibitem{35}
King, A. R., Wynn, G. A., MNRAS, \textbf{310}, 203 (1999) 

\bibitem{36}
Kozhevnikov, V. P., MNRAS, \textbf{422}, 1518 (2012) 

\bibitem{37}
Mason, K.O., MNRAS, \textbf{285}, 493 (1997) 

\bibitem{38}
Masetti, N., Nucita, A. A., Parisi, P.,  A\&A, \textbf{544}, A114 (2012) 

\bibitem{39}
Middleton, M. J., Cackett, E. M., Shaw, C., Ramsay, G., Roberts, T. P., Wheatley, P. J., MNRAS, \textbf{419}, 336 (2012)

\bibitem{40}
Mhlahlo, N., Buckley, D.A.H., Dhillon, V.S., Potter, S.B., Warner, B., Woudt, P.A., MNRAS, \textbf{378}, 211 (2007)

\bibitem{41}
Neustroev, V. V., Tovmassian, G. H., Zharikov, S. V., Sjoberg, G., MNRAS, \textbf{432}, 2596 (2013)

\bibitem{42}
Norton, A.J., Hellier, C., Beardmore, A.P., Wheatley, P.J., Osborne, J.P., Taylor, P., MNRAS, \textbf{289}, 362 (1997)

\bibitem{43}
Norton, A. J., Watson, M. G., King, A. R., Lehto, H. J., McHardy, I. M., MNRAS, \textbf{254}, 705 (1992) 

\bibitem{44}
Oruru, B., Meintjes, P. J., MNRAS, \textbf{421}, 1557 (2012) 

\bibitem{45}
Patterson, J. et al., PASP, \textbf{110}, 415 (1998)  

\bibitem{46}
Patterson, J. et al., PASP, \textbf{123}, 130 (2011) 

\bibitem{47}
Piirola, V., Vornanen, T., Berdyugin, A., Coyne, G.V., ApJ, \textbf{684}, 558 (2008) 

\bibitem{48}
Potter, S. B., Romero-Colmenero, E., Watson, C. A., Buckley, D. A. H., Phillips, A., MNRAS, \textbf{348}, 316 (2004) 

\bibitem{49}
Rodrigues, C. V., Jablonski, F. J., D'Amico, F., Cieslinski, D., Steiner, J. E., Diaz, M. P., Hickel, G. R., MNRAS, \textbf{369}, 1972 (2006)

\bibitem{50}
Scaringi, S. \etal\ A\&A, \textbf{530}, A6 (2011) 

\bibitem{51}
Schwarz, R., Schwope, A. D., Staude, A., Remillard, R. A., A\&A, \textbf{444}, 213 (2005)

\bibitem{52}
Schwope, A. D., Catal\'an, M. S., Beuermann, K., Metzner, A., Smith, R. C., Steeghs, D., MNRAS,  \textbf{313}, 533 (2000)

\bibitem{53}
Schwope, A. D., Schwarz, R., Sirk, M., Howell, S. B., A\&A, \textbf{375}, 419 (2001)

\bibitem{54}
Schwope, A. D., Thomas, H.-C., Mante, K.-H., Haefner, R., Staude, A., A\&A, \textbf{402}, 201 (2003) 

\bibitem{55}
Siegel, N., Reinsch, K., Beuermann, K., van der Woerd, H., Wolff, E., A\&A, \textbf{225}, 97 (1989)

\bibitem{56}
Silva, K. M. G., Rodrigues, C. V., Costa, J. E. R., de Souza, C. A., Cieslinski, D., Hickel, G. R., MNRAS, \textbf{432}, 1587 (2013) 

\bibitem{57}
Thomas, H.-C., Beuermann, K., Reinsch, K., Schwope, A. D., Burwitz, V.,  A\&A, \textbf{546}, A104 (2012)

\bibitem{58}
Vrielmann, S., Ness, J.-U., Schmitt, J. H. M. M., A\&A, \textbf{439}, 287 (2005) 
\bibitem{59}
Warner, B., {\it Cataclysmic variable stars}, (CUP, Cambridge 1995)

\bibitem{60}
Warner, B., PASP, \textbf{116}, 115 (2004)

\bibitem{61}
Warner, B., Pretorius, M. L., MNRAS, \textbf{383}, 1469 (2008) 

\bibitem{62}
Warner, B., Woudt, P. A., MNRAS, \textbf{367}, 1562 (2006) 

\bibitem{63}
Warner, B., Woudt, P. A., MNRAS, \textbf{397}, 979 (2009)

\bibitem{64}
Watson, M.G., King, A.R., Osborne, J., MNRAS, \textbf{212}, 917 (1985)  

\bibitem{65}
Williams G., PASP, \textbf{115}, 618 (2003)

\bibitem{66}
Woudt, P. A., Warner, B., MNRAS, \textbf{400}, 835 (2009)

\bibitem{67}
Woudt, P. A. et al., MNRAS, \textbf{427}, 1004 (2012)

\bibitem{68}
Wynn, G. A., King, A. R., MNRAS, \textbf{275}, 9 (1995) 

\bibitem{69}
Wynn, G. A., King, A. R., Horne, K., MNRAS, \textbf{286}, 436 (1997) 

\end{thebibliography}
\end{document}